# Super-Resolution and Denoising of Corneal B-Scan OCT Imaging Using Diffusion Model Plug-and-Play Priors

Yaning Wang*[a], Jinglun Yu[a], Wenhan Guo[a,b], Ziyi Huang[a], Rosalinda Xiong[a], Yu Sun[a,b], and Jin U. Kang[a]

[a]Department of Electrical and Computer Engineering, Johns Hopkins University, Baltimore, USA;
[b]Data Science and AI Institute, Johns Hopkins University, Baltimore, USA

## ABSTRACT

Optical coherence tomography (OCT) is pivotal in corneal imaging for both surgical planning and diagnosis. However, high-speed acquisitions often degrade spatial resolution and increase speckle noise, posing challenges for accurate interpretation. We propose an advanced super-resolution framework leveraging diffusion model plug-and-play (PnP) priors to achieve 4× spatial resolution enhancement alongside effective denoising of OCT B-scan images. Our approach formulates reconstruction as a principled Bayesian inverse problem, combining Markov chain Monte Carlo sampling with pretrained generative priors to enforce anatomical consistency. We comprehensively validate the framework across multiple domains including *in vivo* fisheye corneal datasets to assess robustness and scalability under diverse acquisition protocols. Comparative experiments against bicubic interpolation, conventional supervised

U-Net baselines, and alternative diffusion priors demonstrate that our method consistently yields more precise anatomical structures, improved delineation of corneal layers, and superior noise suppression. Quantitative results show state-of-the-art performance in peak signal-to-noise ratio, structural similarity index, and perceptual metrics, confirming the framework's efficacy across species and imaging conditions. This work highlights the potential of diffusion-driven plug-and-play reconstruction to deliver high-fidelity, high-resolution OCT imaging, supporting more reliable clinical assessments and enabling advanced image-guided interventions. Our findings suggest the approach can be extended to other biomedical imaging modalities requiring robust super-resolution and denoising.

## 1. INTRODUCTION

Optical coherence tomography (OCT) is rapidly transitioning from a diagnostic modality into an intraoperative imaging tool, enabling visualization of fine tissue microstructures and real-time surgical guidance[1–6]. Ophthalmology remains its most prominent clinical domain. A major challenge in ophthalmic OCT is involuntary eye motion during acquisition, including tremors, drifts, micro-saccades, and axial shifts induced by vascular pulsation and respiration[7]. These motions distort sampled voxels, degrade image quality, and compromise reproducibility of quantitative measurements, especially in clinical environments with uncooperative patients.

Both hardware- and software-based motion mitigation strategies have been explored. Hardware solutions often require auxiliary sensing hardware to track ocular movement, resulting in increased complexity and cost[8,9]. Classical software approaches rely on hand-crafted registration and motion models, but accuracy remains limited under unpredictable and non-rigid tissue motion[10,11].

Deep learning (DL) methods have shown promise for motion removal and image super-resolution (ISR) in OCT. ISR seeks to reconstruct high-resolution (HR) images from degraded low-resolution (LR) measurements corrupted by blur, noise, and motion. However, supervised DL models trained on synthetic LR–HR pairs typically assume simplified motion (often rigid), struggle to generalize to real degradation, and produce over-smoothed reconstructions that lack structural fidelity[12].

Recent efforts introduce generative priors through generative adversarial networks (GANs) or sparsity-based residual models to improve perceptual realism[13]. Yet GANs suffer from training instability and limited mode coverage, while sparsity priors alone cannot fully capture complex texture statistics. More recently, diffusion models (DMs) have emerged as powerful generative priors with strong perceptual and structural modeling capabilities, demonstrating state-of-the-art results in real-world restoration tasks[14].

*ywang511@jhu.edu

In this work, we propose an iterative plug-and-play diffusion model (PnP-DM) framework for high-fidelity OCT reconstruction from sparsely sampled corneal B-scans. We formulate reconstruction as an inverse problem and alternate between a likelihood-based data-consistency step and a prior refinement step based on a pre-trained diffusion model. To obtain realistic LR–HR training pairs, we acquire under sampled high-speed scans and apply a DL-based up sampling pipeline to synthesize HR targets. Integrated within a Markov chain Monte Carlo (MCMC) sampling procedure, PnP-DM enables efficient posterior sampling that balances prior strength and measurement fidelity. Evaluations on *in vivo* datasets demonstrate improved structural preservation and perceptual quality compared to a strong 2D-UNet baseline.

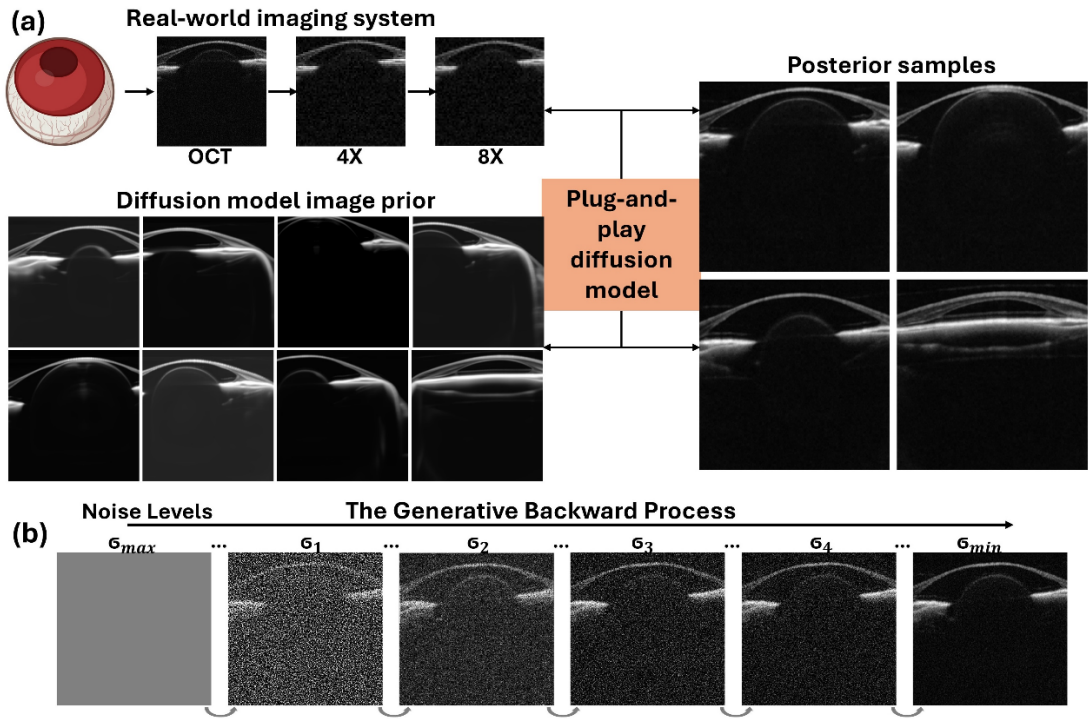

Figure 1. (a) Overview of the proposed PnP-DM framework, which alternates data-consistent likelihood updates with an EDM-based prior trained on *in vivo* fisheye OCT B-scans. (b) The backward diffusion process progressively denoises latent samples, transforming noise into a clean high-resolution OCT reconstruction.

## 2. METHODS

We adopt a plug-and-play diffusion model for OCT super-resolution, following the iterative Bayesian framework introduced in[14]. The framework enables high-fidelity reconstruction across varying under sampling rates. Figure 1 illustrates the proposed pipeline.

**Problem Formulation**

Given a LR measurement $y \epsilon R^m$, the goal is to estimate the HR OCT image $x \epsilon R^n$. The forward model is expressed as

$$y = A(x) + n, \qquad (1)$$

where $A(\cdot): R^n \rightarrow R^m$ is a linear operator and *n* denotes measurement noise.

For super-resolution, the acquisition is modeled as:

$$y \sim N(P_f x, \sigma_y^2 I), \qquad (2)$$

where $P_f \epsilon R^{n_f \times n}$ performs block-averaging down sampling with factor $f = 4$. The singular value decomposition (SVD) of $P_f$ is precomputed[15] for efficient matrix operations.

We formulate SR as posterior sampling combining a likelihood term and a diffusion prior. The method alternates between:

- Likelihood step: data-consistency update,
- Prior step: diffusion-based denoising.

This follows the Split Gibbs Sampler (SGS) framework[16].

**Likelihood Step**

At iteration $q$, the latent variable $z_{(q)}$ is sampled as:

$$z_{(q)} \sim \pi^{Z|X=x^{(q)}}(z) \propto exp\left(-f(z;y) - \frac{1}{2\rho^2} \| x^{(q)} - z \|_2^2 \right), \tag{3}$$

where $f(z; y)$ is the data-fidelity term. For Gaussian noise:

$$f(x; y) = \frac{1}{2} \| y - Ax \|_\Sigma^2, \tag{4}$$

The resulting conditional distribution is Gaussian:

$$\pi^{Z|X=x} = N(m(x), \Lambda^{-1}), \tag{5}$$

where,

$$\Lambda = A^T \Sigma^{-1} A + \frac{1}{\rho^2} I, m(x) = \Lambda^{-1}\left(A^T \Sigma^{-1} y + \frac{1}{\rho^2} x\right). \tag{6}$$

Matrix operations exploit the SVD of $A$ for computational efficiency.

**Prior Step: EDM Diffusion Refinement**

To regularize reconstructions, we adopt a diffusion prior using the EDM framework[17], which unifies major diffusion formulations. The reverse-time stochastic differential equation (SDE) is:

$$dx_t = g(x_t, t)dt + \sigma(t)dw_t, \tag{7}$$

where $\sigma(t)$ is the noise schedule and $w_t$ is a Wiener process. Solving this SDE backward denoises samples toward the clean image manifold.

**PnP-DM Sampling with Annealing**

The alternating likelihood–prior updates are controlled by a coupling parameter $\rho_q$. We adopt an exponential decay schedule:

$$\rho_0 = 10, \rho_{min} = 0.3, \rho_q = \alpha^q \rho_0, \alpha = 0.9 \tag{8}$$

which accelerates mixing and stabilizes convergence under highly ill-posed super-resolution conditions.

## 3. RESULTS

We evaluate the proposed PnP-DM framework on *in vivo* fisheye OCT datasets to assess reconstruction fidelity under sparse sampling. Experiments focus on structural preservation, noise suppression, and perceptual realism during 4× super-resolution from 64 × 64 to 256 × 256 B-scans.

**Datasets and Preprocessing**

The training dataset consists of 164 *in vivo* fisheye volumes with dimensions 1024 × 1024 × 128. All scans are normalized to the range [0, 1]. For multi-resolution training, high-resolution slices (N = 20,992) are downsampled by a factor of 4 along the depth and fast-scan axes to synthesize paired LR measurements. Validation and testing are performed on *in vivo* fish corneal datasets acquired using conventional scanning protocols.

**Implementation Details**

All experiments are implemented in PyTorch and executed on NVIDIA A100 GPUs. Models are trained for 2,000 epochs using the Adam optimizer with a learning rate of $10^{-5}$ and exponential decay of 0.9. Inputs use FP16 precision and are resized to 256 × 256. The pretrained VP diffusion score network is converted to EDM form using VP preconditioning. During inference, each under sampled B-scan undergoes 100 PnP-DM iterations, and each Markov chain requires approximately 80 seconds to converge.

**Compared Methods and Metrics**

We benchmark the proposed approach against:

- Bicubic interpolation,
- 2D-UNet baseline,
- PnP-DM variants with VP[18], VE[19], DDPM[18], and EDM[17] diffusion priors.

Reconstruction performance is quantified using peak signal-to-noise ratio (PSNR), structural similarity index measure (SSIM), and learned perceptual image patch similarity (LPIPS). For each diffusion-based variant, 100 posterior samples are generated and quantitative metrics are reported using their mean reconstruction.

**Qualitative Comparisons**

Figure 2 illustrates representative super-resolution results. Bicubic interpolation produces oversmoothed structures and blurred boundaries, particularly in low-SNR regions. The 2D-UNet baseline suppresses speckle, but overly smooths textures and attenuates anatomical boundaries. In contrast, all PnP-DM variants yield sharper structural details, well-preserved stromal and iris interfaces, and improved speckle suppression. Reconstructions from different diffusion priors exhibit comparable perceptual quality, demonstrating the robustness of the plug-and-play sampling process.

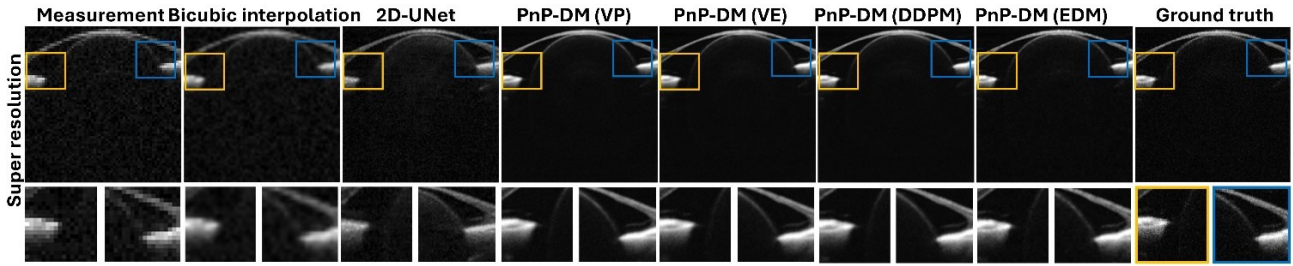

Figure 2. Visual examples for the super resolution problem

**Quantitative Comparisons**

Table 1 summarizes the performance on 100 fisheye OCT images. PnP-DM significantly improves over classical interpolation and the 2D-UNet baseline across all metrics. VE and EDM diffusion priors achieve the highest PSNR/SSIM and lowest LPIPS values, indicating superior alignment of the learned prior with the OCT forward model.

Overall, the results confirm that PnP-DM effectively balances measurement consistency and structural priors, producing high-quality super-resolution outputs from sparse OCT data and demonstrating strong potential for integration into high-speed clinical OCT workflows.

Table 1: Quantitative evaluation on super-resolution OCT for 100 fisheye corneal OCT images, **Bold**: best.

|  | BI | 2D-UNet | PnP-DM (VP) | PnP-DM (VE) | PnP-DM (DDPM) | PnP-DM (EDM) |
|---|---|---|---|---|---|---|
| PSNR↑ | 28.13 | 30.07 | 30.48 | **32.50** | 32.37 | 32.14 |
| SSIM↑ | 0.4607 | 0.4863 | 0.7163 | **0.7224** | 0.7214 | 0.7150 |
| LPIPS↓ | 0.4024 | 0.2072 | 0.1415 | 0.1257 | 0.1264 | **0.1201** |

## 4. CONCLUSION

We presented a super-resolution OCT reconstruction framework based on plug-and-play diffusion priors. By formulating reconstruction as an inverse problem and alternating between diffusion-based prior refinement and likelihood-driven data consistency within an MCMC sampling procedure, the pro- posed PnP-DM enables efficient and stable inference from sparse OCT measurements. Experiments on *in vivo* fish-eye datasets show that PnP-DM consistently outperforms a strong UNet baseline, yielding sharper structural detail and improved speckle suppression. These results demonstrate the promise of diffusion-driven priors for high-speed, high-resolution OCT imaging and motivate further integration into clinical workflows.

## ACKNOWLEDGEMENTS

This work was supported by National Institute of Health Grant Award No. 1R01EY032127 (PI: Jin U. Kang), the study was conducted at Johns Hopkins University.
We thank Dr. Ruizhi Zuo for his assistance with data collection.